# Radar Range Deception with Time-Modulated Scatterers

V. Kozlov[1], D. Vovchuk[1,2] and P. Ginzburg[1]

[1]School of Electrical Engineering, Tel Aviv University, Tel Aviv, 69978, Israel
[2]Department of Radio Engineering and Information Security, Yuriy Fedkovych Chernivtsi National University, Chernivtsi, 58000, Ukraine

*Abstract* – **Modern radar systems are designed to have high Doppler tolerance to detect fast-moving targets. This means range and Doppler estimations are inevitably coupled, opening pathways to concealing objects by imprinting artificial Doppler signatures on the reflected echoes. Proper temporal control of the backscattered phase can cause the investigating radar to estimate wrong range and velocity, thus cloaking the real position and trajectory of the scatterer. This deception method is exploited here theoretically for arbitrary Doppler tolerant waveforms and then tested experimentally on an example of the linear frequency modulated radar, which is the most common waveform of that class used in practice. The method allows retaining radio silence with a semi passive (battery assisted) approach that can work well with time-dependent metasurfaces. Furthermore, as an insight into new capabilities, we demonstrate that temporally concealed objects could even be made to appear closer than they truly are without violating the laws of relativity.**

## I. INTRODUCTION

Contemporary radar systems play a significant role in numerous sensing applications [1]–[7] and are likely to remain just as relevant in the foreseeable future, both as standalone platforms as well as parts of a fused sensory network. The key advantage of such systems lies in their relatively low operational frequency band in the electromagnetic spectrum, allowing interrogation through fog and other obstacles that render optical investigation difficult. The vital importance of such systems, as well as their analogues in sonar and LIDAR, birthed entire scientific and technological branches, ushering advances in both hardware and software solutions [8]–[12]. The success and proliferation of radar systems was inevitably met with a relentless pursuit for electronic countermeasures (ECM) to avoid detection, which were themselves soon followed by counter countermeasures (ECCM) and so on [13]–[18]. Stealth technologies were introduced to minimize the signatures measured by the receivers [19], [20], offering a passive solution to the problem of evading detection without transmitting electromagnetic radiation, which could expose the radar to honing by a sophisticated ECCM system. Yet no matter how absorbent the material is, no matter how careful the geometry is designed, the standoff distance could only be reduced by a factor. Moreover, multi-static radars and interrogation by discerning reflection from upper layers of the atmosphere can challenge stealth technologies quite significantly. To further confuse the investigating radar, jamming measures were soon added. By actively transmitting noise towards the radar, its dynamic range could be reduced, and detection probability diminished at the cost of foregoing radio silence. Spoofing methods followed, with the most rudimentary solutions being the release of chaff to decoy the radar [21]–[24], and the more advanced were the introduction of repeaters that could imprint careful signatures on the reflected echoes, causing the radar to deduce the wrong trajectory and location of the targets of interest, producing so called ghosts [25]–[28]. The drawback of such methods was rooted in their failure to suppress the echo from the target itself, relying instead on amplifying the spoofed echoes from the repeater, in an attempt to draw the radar to track the larger return rather than the object of interest. It did not take long for spoofing ECM to be met with smart ECCM signal processing [29]–[31], severely challenging its effectiveness. In recent years the emerging field of metamaterials and metasurfaces had rapidly developed [32]–[34], enabling novel types of passive stealth capabilities not possible before [35]–[37].

Today, dynamic control over metamaterial scattering properties is a rapidly developing field. By carefully controlling the reflection coefficient of the scatterer in time, it is possible to imprint arbitrary signatures on the backscattered echoes [38]–[42]. The goal of such meta-covers is achieving the advantages of state-of-the-art repeaters without suffering from the drawback of merely superimposing on top of the reflection from the real target. The concept of using metamaterial surface covers is illustrated in Fig.1, where a drone (as an illustrative example) is depicted with and without a time-dependent cloak while it is observed by the same radar system. The concealed drone in Fig1(b) successfully modulates the reflected echoes in such a way that it appears closer than it really is. This statement might at first appear to contradict intuition, seeing as a pulse cannot travel faster than light. This apparent 'faster than light travel' can only be achieved for a certain, yet widely used class of radar signals, as will be discussed ahead.

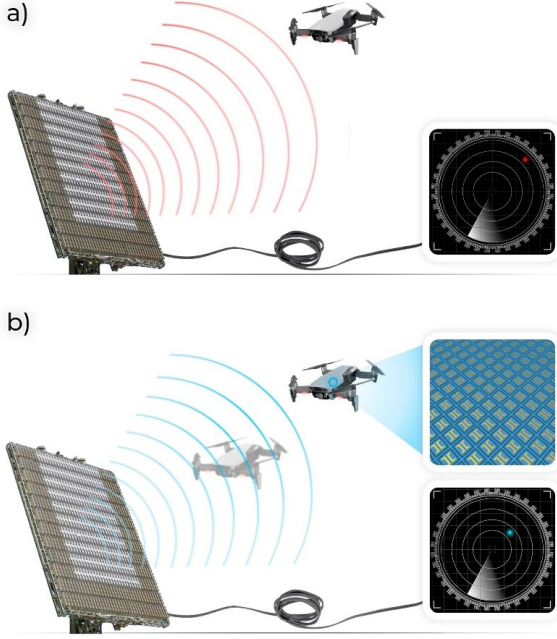

**Figure 1**: Illustration of the proposed deception concept. A radar observing a drone with and without metasurface cover (a) The drone without cover is correctly detected. (b) Dynamic control of the scattering properties from the drone conceals its true location, making it appear closer (or further apart, on demand) than it really is to the investigating radar.

The manuscript is organized in the following way. First, a theoretical derivation of the proposed deception concept and how to counteract it is presented for a general class of Doppler tolerant waveforms. Then an experiment is conducted in the anechoic chamber, showing that it is indeed possible to manipulate a frequency modulated continuous waveform (FMCW) radar to conclude arbitrary position of the scatterer simply by dynamic control over the scattered phase. The 'Outlook and conclusion' section follows.

## II. THEORY - RANGE DECEPTION OF DOPPLER TOLERANT RADARS

Radar systems rely on signal processing to extract information about the observed environment from reflected echoes. To achieve this, the transmitted waveform needs to be carefully selected to meet the specifications. The parameters of interest are, among others, range and Doppler resolutions [43]–[45], which are preferably as small as possible. These metrics of a waveform $s(t)$ can be conveniently viewed through the narrowband ambiguity function (wideband signals are beyond the scope of this manuscript), given by [43]:

$$\chi_{S(t)}(\tau, \omega_D) = \int_{-\infty}^{\infty} e^{-i\omega_D t} s(t) s^*(t-\tau) dt. \quad (1)$$

This is a function of the Doppler frequency $\omega_D$ due to the motion of the reflecting point-like target, located at a distance $R$, which translates to delay $\tau = \frac{2R}{c}$ through the speed of light $c$. The function in Eq.1 describes the output of the matched filter for a specific waveform, which ensures maximal signal to noise ratio at the range of the target. In other words, the matched filter is searching for the signal in the sampled data by correlating it with the known transmitted waveform, while the ambiguity function also considers the output due to possible distortion of the returning echo by the Doppler effect. Consider three commonly used waveforms with compact support on (0,T) [43]:

$$s_{Noise}(t) = n_\sigma(t) u\left(\frac{t}{T}\right), \quad (2)$$

$$s_{LFM}(t) = e^{-i\left(\omega_0 + \frac{\Delta_\omega}{2T}t\right)t} u\left(\frac{t}{T}\right), \quad (3)$$

$$s_{P_3}(t) = e^{-i\varphi_M(t)} u\left(\frac{t}{T}\right), \quad (4)$$

where $u(t) = \begin{cases} 1 & 0 < t < 1 \\ 0 & o.w \end{cases}$ is the rectangular pulse function, $n_\sigma(t)$ is a zero-mean complex Gaussian random process with standard deviation $\sigma$, $\omega_0$ is the initial frequency of the linear frequency modulated (LFM) chirp, which can be positive (up chirp) or negative (down chirp), and $\Delta_\omega$ is the bandwidth of the chirp, i.e. the maximal deviation from the initial frequency. The polyphase waveform $P_3$ in Eq.4 can be thought of as an M-sampled variation of the LFM, with the phases $\varphi_M(t) = \sum_{m=1}^{M} \varphi_m u\left(\frac{t - \frac{T}{M}(m-1)}{\frac{T}{M}}\right)$ and $\varphi_m = \begin{cases} (m-1)^2 & m \text{ even} \\ (m-1)m & m \text{ odd} \end{cases}$. This is the place to note that the complex analytical representations of the signals is used throughout the manuscript.

Fig.2 shows the ambiguity function of the above waveforms for arbitrarily selected parameters, where the pulse width is $T = 50\mu s$, the sample rate is 5MHz and the pulse repetition frequency (PRF) is 10kHz. The axis are normalized to present fractional Doppler to bandwidth ratio ($\frac{\omega_D}{\Delta_\omega}$) and range to unambiguous range ratio ($\frac{R}{R_{unambiguous}}$) for illustrative purposes, as will be discussed ahead. Note that positive Doppler frequencies represent approaching targets while negative Doppler frequencies represent receding targets. Fig.2(a) shows the results for the noise waveform in Eq.2 with $\sigma = 1$, having a very well-defined peak at the origin, often called a 'thumbtack', which achieves excellent range and

Doppler resolutions (recall that the cross-correlation of white noise is a delta function, which is appealing for obtaining superior range resolution). However, the noise waveform suffers immensely from Doppler intolerance, which means that if the reflecting target was moving, the output of the matched filter would be a horizontal cut above or below the origin, where no ambiguity volume is left, rendering the target practically invisible. To overcome this issue, it is better to have a 'ridge-like' autocorrelation function, such as illustrated in Fig.2(b) and Fig.2(d), showing the ambiguity function of the LFM waveform in Eq.3 with in an up and down chirp respectively of $|\Delta_\omega| = 2\pi \times 0.5 MHz$ and $\omega_0 = 0$ for simplicity. This is the most commonly used Doppler tolerant waveform at present day and it will be used for the experiment ahead. This waveform clearly has a significant output even for a moving target, as can be seen by taking horizontal cuts of the plots. While the term 'Doppler tolerance' is somewhat open to interpretation [46]–[48], it can be said that any waveform that can still produce a strong output for a significant band of Doppler frequencies is tolerant. Another example of such a waveform appears on Fig.2(c), where the phase coded $P_3$ waveform in Eq.4 with $M = 25$ is shown. It bears a lot of resemblance to the down chirped LFM, but with visible gaps, meaning that for certain target velocities detection might not be possible. Another apparent result, which can be deduced from the symmetries of the ambiguity function, is that Doppler tolerant waveforms inevitably couple range with Doppler. This means that any moving target would produce an error in range which in these example is proportional to the radial velocity of the observed scatterer [49], [50]. This relationship can be exploited by artificially imprinting Doppler signatures on the echoes reflected from the target, as will be shown experimentally ahead. Notice that it is possible to deceive the radar into concluding that the target is even closer than it truly is without violation of relativity, but simply by the virtue of the properties of the matched filter detection method for this class of waveforms. The axes of Fig.2 are stressing the fact that it is possible to create a significant error in range (positive or negative, at will), reaching as high as 20% from the maximal unambiguous range ($R_{unambiguous} = \frac{c}{2PRF}$), given the parameters set above. It is also evident that such artificial modulation would not lead to the signal being completely filtered out by the bandpass filter of the receiver, since it is well within the required operational bandwidth of the system. By reducing the PRF and increasing the bandwidth of the chirp $\Delta_\omega$, the attainable range shift will diminish, and the proposed deception method will be more limited.

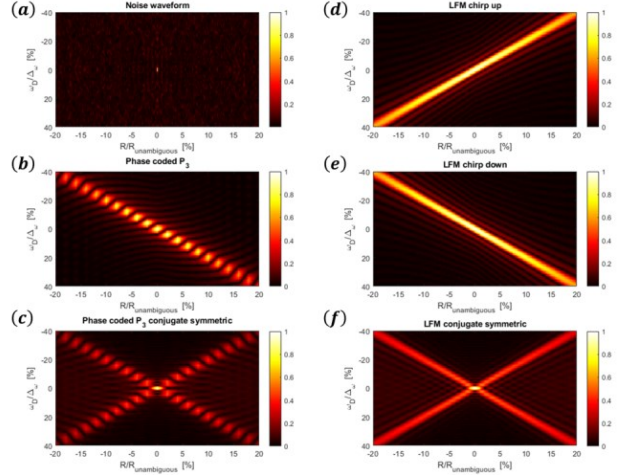

**Figure 2**: Absolute values of the ambiguity function of six different waveforms as a function of fractional Doppler to bandwidth ratio and range to unambiguous range ratio. (a) Noise waveform (b) Phase coded $P_3$ (c) Phase coded $P_3$ after conjugate symmetric appending (d) LFM chirp up (e) LFM Schirp down (f) LFM after conjugate symmetric appending (chirp up then back down).

### III. COUNTERMEASURES - CONJUGATE SYMMETRIC WAVEFORMS

In order to counter spoofing attempts of this kind, diversity can be introduced to the transmitted waveform. For example, it is possible to alternate between an up chirp and a down chirp of the LFM, producing two different alternating ranges for the target that can be resolved as the mean of the two. Another approach, which to the best of our knowledge is first suggested here, is applicable to all waveforms. By appending any signal $s(t)$ with compact support on (0,T) to its time reversed and conjugated self, we arrive at the new conjugate symmetric (CS) waveform:

$$s_{CS}(t) = s(t) + s^*(-t + 2T), \qquad (5)$$

which is of length $T_{CS} = 2T$ and has the symmetric property around its center $s_{CS}\left(\frac{T_{CS}}{2} - t\right) = s_{CS}^*\left(\frac{T_{CS}}{2} + t\right)$. It is straightforward to show by substitution to Eq.1 that the ambiguity function of these special waveforms is symmetric in respect to both axis:

$$\begin{aligned}|\chi_{S_{CS}(t)}(\tau, \omega)| &= |\chi_{S_{CS}(t)}(-\tau, \omega)| = \\ |\chi_{S_{CS}(t)}(\tau, -\omega)| &= |\chi_{S_{CS}(t)}(-\tau, -\omega)|.\end{aligned} \qquad (6)$$

The result of applying the CS process to the $P_3$ and LFM waveforms is shown on Fig.2(c) and Fig.2(f) respectively, where the symmetry of the ambiguity function is apparent. Such waveforms would produce two separate peaks in the autocorrelation (horizontal cut of the ambiguity function) for a moving target, with the true location always being located in between the two peaks, allowing for correct range detection. The suggested CS process can serve to retain a lot of the advantages of the desired waveform, while also making sure it is hard to spoof by artificial Doppler modulation. In fact, the result is well known for the private case of FMCW radars, which implement the matched filter in hardware by mixing the received signal with the transmitted chirp, as will be done in the experiment ahead. To detect the Doppler of the target, a triangular frequency modulation can be used with an up chirp followed by a down chirp, producing two peaks in the spectrum of the mixer's output, with the distance between the peaks proportional to the velocity of the target, and its location proportional to the mean of the peaks. The CS process suggested here expands this result to all Doppler tolerant waveforms.

## IV. EXPERIMENTAL VERIFICATION – RANGE DECEPTION OF FMCW RADARS

To validate the theoretical results described above, an experiment was conducted in the anechoic chamber as shown on Fig.3. An FMCW radar was constructed by up-converting a 40MHz linear down-chirp with a period of $50\mu s$, produced by an AFG-3051, to the central frequency of 750MHz. The radar used a circulator for isolation of the transmitting and receiving channels, which were both fed by the same Yagi-Uda antenna. The time-modulated scatterer that is shown on Fig.3 was assembled from a pair of identical antennas, which were connected with a long delay line made out of cables to simulate a more distant target. A phase shifter was placed along the connecting line as seen in the inset to Fig.3. The phase shifter was dynamically controlled by a biasing voltage applied to a vector modulator (AD8340), capable of providing an arbitrary time-dependent phase profile. An amplifying stage was added following the phase shifter in order to compensated for the losses in the cables.

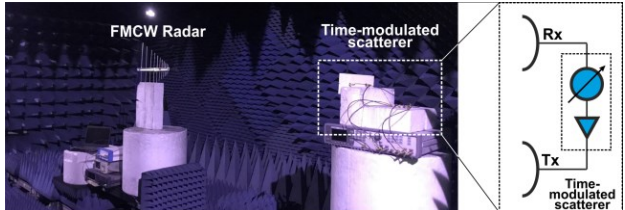

**Figure 2**: The experimental set up in the anechoic chamber, an FMCW radar is directed at a time-modulated scatterer. The schematic inset shows the electronics of the scatterer, allowing fully dynamic reflected phase control.

The reflected signal from the time-modulated scatterer was picked up by the radar's antenna and was continuously mixed with the transmitted chirp to produce the baseband output. The baseband was low pass filtered before being observed in the frequency domain using a sampling scope, as can be seen on the insets in Fig.4. The signal at the baseband can be computed by using Eq.3 and solving:

$$s_{baseband}(t) = Re\{s_{LFM}(t)\}Re\{s_{LFM}(t-\tau)e^{-i\omega_D t}\} \approx Cos\left(\underbrace{\left(\frac{\Delta_\omega \tau}{T}+\omega_D\right)t}_{\omega_{baseband}}\right)u\left(\frac{t-\tau}{T-\tau}\right), \quad (7)$$

where we have neglected the second harmonic of the carrier, which is outside of the bandwidth of the filter, as well as the phases and contributions of the mixing between the reflected echo with the next transmitted chirp (assuming $T \gg \tau$). The use of the real operator is required before multiplication in Eq.7 seeing as mixing is a non-linear operation. The resulting dominant frequency at the baseband $\omega_{baseband}$ is therefore a function of the delay and Doppler, and its measurement allows calculating the range of the target:

$$R = \frac{cT}{2|\Delta_\omega|}(\omega_{baseband} - sign(\Delta_\omega)\omega_D), \quad (8)$$

where $sign(x) = \begin{cases} +1 & x > 0 \\ -1 & x < 0 \end{cases}$ and we assumed without loss of generality that $\left|\frac{\Delta_\omega \tau}{T}\right| < |\omega_D|$. Eq.8 clearly reveals the coupling between range and Doppler that was explored in the theoretical section through the prism of the ambiguity function, where positive Doppler shifts (approaching targets) represent upconversion of the reflected chirp, and negative shifts (receding targets) represent downconversion of the chirp.

The phase shifter of the time-modulated scatterer was configured to imprint fake Doppler signatures, on the reflected echoes, while the corresponding distance of the target was deduced from measuring the dominant frequency at the baseband of the radar receiver (see insets to Fig.4) and using Eq.8 with $\omega_D = 0$. The results are plotted on Fig.4, where a linear relation can be observed between the modulation frequency, which corresponds to the fake Doppler shift frequency, and the estimated range of the target. This linear relationship is in very good agreement with the results derived in Eq.8 and Fig.2, demonstrating full control over the perceived location of the target, and even making it possible to create the illusion that the target is closer than it really is. The insets on Fig.4 show the baseband signals recorded by the sampling scope at each of the corresponding modulation frequencies. Some additional peaks can be observed, attributed in small part to multipath interference and the neglection of some terms in the derivation leading to

Eq.7. The most prominent source however is the formation of the so called "picket fence" that arises whenever a periodic signal is sampled, forming an equidistant comb around the dominant peak, sometimes even obscuring it. The root cause is related to the periodicity of the phase modulation and the continuous transmitted radar waveforms.

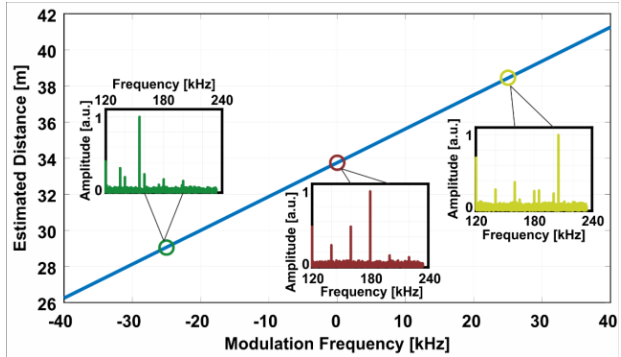

**Figure 3**: Demonstration of full control over the perceived range of the static time-modulated scatterer. By controlling the modulation frequency of the reflected phase from the object, it may appear further and even closer than it truly is. Insets show the baseband signal at the FMCW radar for selected modulation frequencies.

## V. OUTLOOK AND CONCLUSION

A temporal phase modulation approach for radar deception was demonstrated. In contrast with existing passive (e.g., stealth and metamaterials) and active (e.g., jamming, spoofing) approaches, time-modulation of the reflected echoes does not require neither complex shaping of enclosures around a target, nor expensive and cumbersome power consuming electronics. Here we introduced a deception concept, which is based on traversing "journeys" along the ambiguity function, which characterizes the radar's measurement accuracy, taking advantage of the signal processing weaknesses that are inherent to a very broad class of radar systems. We have demonstrated that Doppler tolerant waveforms are susceptible to range deception by exploiting one of their greatest strengths, but also provided a solution to counter these shortcomings with a simple process of conjugately symmetrizing the transmitted waveforms. In particular, we showed experimentally that FMCW radars can interpret targets with time-modulated scattering cross sections as if they were closer than they truly are. While this behavior might at first glance contradict the laws of relativity, a reasonable objection had the target been interrogated by a short pulse, this is shown to be perfectly normal in the context of Doppler tolerant waveforms that rely on matched filters, paving the way for temporal-phase deception strategies. As an outlook, the proposed concept can be used to conceal large targets. In this case, time-modulated metasurfaces, capable of providing dynamic $2\pi$ control over the phase, can be used to cover scattering centers of interest. Another application concerns smart electromagnetic chaff, which can be used for creating fake radar targets that correctly mimic the expected Doppler velocity. While classical "static chaff" is straightforwardly filtered out with clutter filters, smart time-modulated structures are immune against those countermeasures.

## ACKNOWLEDGMENTS

The research was supported in part by Department of the Navy, Office of Naval Research Global under ONRG award No. N62909– 21–1–2038.